\documentclass [12pt]{article}
\usepackage{setspace}
\usepackage{amsmath}
\usepackage{comment}
\usepackage[ruled,vlined]{algorithm2e}
\usepackage{enumitem, booktabs}
\usepackage{rotating}
\usepackage{threeparttable, tablefootnote}
\usepackage{makecell}
\usepackage{natbib}
\newlist{steps}{enumerate}{1}
\setlist[steps, 1]{label = Step \arabic*:}
\title{Modeling Event Dynamics by Self-Exciting Processes with Random Memory}
\author{K. Ken Peng, X.\ Joan Hu and Tim B.\ Swartz
\thanks{
K. Ken\ Peng is a Postdoctoral Fellow in the Department of Civil Engineering,
University of Ottawa, and the Department of Statistics and Actuarial Science,
Simon Fraser University.
X.\ Joan\ Hu and Tim B.\ Swartz are Professors of Statistics in the Department of
Statistics and Actuarial Science, Simon Fraser University, 8888 University Drive,
Burnaby, BC, Canada V5A~1S6.
Corresponding author: Tim B.\ Swartz.
}}

\date{ }
\begin{document}
\maketitle

\begin{abstract}
\noindent
Event history data from sports competitions have recently drawn increasing attention in sports analytics to generate data-driven strategies. Such data often exhibit self-excitation in the event occurrence and dependence within event clusters.
The conventional event models based on gap times may struggle to capture those features. In particular, while consecutive events may occur within a short timeframe, the self-excitation effect caused by previous events is often transient and continues for a period of uncertain time. This paper introduces an extended Hawkes process model with random self-excitation duration to formulate the dynamics of event occurrence. We present examples of the proposed model and procedures for estimating the associated model parameters. We employ the collection of the corner kicks in the games of the 2019 regular season of the Chinese Super League to motivate and illustrate the modeling and its usefulness. We also design algorithms for simulating the event process under proposed models. The proposed approach can
be adapted with little modification in many other research fields such as Criminology and Infectious Disease.

\end{abstract}

\vspace{3mm}

\clearpage
\section{Introduction}
\label{sec:intro}
First introduced by \citet{hawkes1971spectra}, 
the Hawkes process is a class of point process models where the intensity function of event occurrence depends on the past events and all the past events {\it self-excite} the on-coming events. The Hawkes process is particularly suitable for characterizing phenomena of event clustering. It has been an important tool for modeling event occurrences across various fields, including earthquakes \citep{ogata1998space, zhuang2002stochastic, kwon2023flexible}, finance \citep{bacry2014hawkes}, and infectious diseases \citep{schoenberg2023estimating, garetto2021time}. 

Numerous extensions of the Hawkes process
have been proposed to adapt to different application scenarios. For instance, \citet{daw2022ephemerally} investigates an ephemeral self-exciting Hawkes process, while \citet{wu2022markov} studies a Markov-modulated Hawkes process to analyze bursty event dynamics in social interactions. \citet{garetto2021time} proposes a time-modulated Hawkes process to model the spread of COVID-19, and \citet{cai2024latent} introduces a framework based on the Hawkes process for learning latent network structures from multivariate event data. Furthermore, \citet{fierro2015hawkes} examines the Hawkes process under varying excitation functions and studies its asymptotic properties.

It is unrealistic in some applications to assume that all past events continuely influence the coming events. This has motivated the development of ``one-memory" Hawkes processes, where the event intensity depends only on the most recent event instead of the entire history. That model is closely related to the ``carryover effect" model in the situations where evidence of short-term influence is sought by examining whether event intensities are temporarily elevated following an event occurrence; see, for example, \citet{cciugcsar2012assessing}. Such effect carryover is often not directly observable and requires careful statistical modeling.

In the field of sports analytics, event data recorded from competitions can reveal the mechanisms behind the event occurrences and provide valuable insights into sports tactics and betting markets. Among these, corner kicks in association football (soccer) are of particular interest since they represent important scoring opportunities and offer insights into set-piece strategies and game dynamics. Thus they are critical for performance analysis and tactical planning. \citet{peng2024} investigates the time elapsed from a natural starting point until a corner kick occurs based on event history data from 233 matches in the 2019 Chinese Super League (CSL) season. A key limitation in \citet{peng2024} is the strong assumption that waiting times for corner kicks are independent conditional on covariates — an assumption that may not always hold. Popular strategies for addressing the potential dependency of the consecutive events are introducing frailty random effects, employing copula models, or formulating the event process as a counting process without independent increcriments.

Another particular complication in analyzing corner kicks is the likelihood of a subsequent corner kick occurring shortly after a previous one. This occurs because a corner kick directed towards the goal area may be immediately deflected by a defending player, go out of bounds and lead to a subsequent corner kick. A natural approach to model this feature in the counting process framework is to employ the Hawkes process, wherein the occurrence of one corner kick increases the chance of another occurring soon after. However, the classic Hawkes process model does not accommodate some distinctive characteristics of soccer corner kicks. First, while some corner kicks indeed trigger quick follow-up kicks, not all do: \citet{peng2024} suggests that only about 10\% of corner kick occurrences form such ``quick" clusters. Second, the influence of a previous corner kick on a subsequent one is constrained within a short and unknown time window. Provided that a significant amount of time has elapsed, it is reasonable to assume the current corner kick is independent of previous ones.

Motivated by those considerations, we introduce an extended Hawkes process model. It formulates the influence by previous events using a partly-hidden semi-Markov process.  Procedures for estimating the model parameters are developed and illustrated by the corner kick data from the CSL. The proposed approach in fact is applicable in many applications where the event processes pose similar features as corner kicks in a soccer game. For example, it
can be used to model event-time data exhibiting heterogeneous risk patterns, such as patient survival times with a treatment-related change point. The rest of this paper is organized as follows. Section \ref{sec:mod} presents the proposed model and associated estimation procedures. Section \ref{sec:analysis} reports empirical studies on the proposed approach, an analysis of the corner kick data from the 2019 CSL season and some simulation experiments. Section \ref{sec:dis} concludes with a summary of the contributions and potential directions for future research.

\section{An Extended Hawkes Process Model}
\label{sec:mod}
\subsection{Notation and Modeling}
\label{sec:nota}

Consider a counting process $\{N(t), t \geq 0\}$, where $N(t)$ represents the number of event occurrences over the time
period $(0,t]$. Let $\mathcal{H}(t)$ denote the history of the process up to and not including time $t$, which is the $\sigma$-field generated by $\{N(u): 0<u<t\}$. Denote the event times over the whole follow-up period $(0, E]$, where $E$ denotes the end of the observation window, by $T_1,\ldots,T_K$ with the total number of events $K = N(E)$ and the potential covariates observed over $(0,E]$ by $\pmb{Z}$.

We propose a novel extension of the Hawkes process model with the conditional intensity $\lambda(t|\mathcal{H}(t), \tau, \pmb{Z})$ switching between two regimes based on a partly hidden semi-Markov structure. Specifically, the intensity at time $t$ is 
\begin{equation}
  \lambda(t|\mathcal{H}(t), \tau, \pmb{Z}) = \begin{cases}
\lambda_1(t|\mathcal{H}(t),\pmb{Z}), & S(t) = 1 \\
\lambda_0(t|\mathcal{H}(t),\pmb{Z}), & S(t) = 0,
\end{cases}
\label{proposed}
\end{equation}
where, with $\tau$ an unobserved random duration, $S(t)$ is an alternating binary, partly hidden process defined as
\begin{equation*}
  S(t) = \begin{cases}
  1, & t-T_{N(t-)} \leq \tau\\
0, & t-T_{N(t-)} > \tau.
\end{cases} 
\end{equation*}
Here $N(t-)$ is the number of preceding event occurrences at time $t$; $T_{N(t-)}$, the time of the most recent occurrence. The binary process $S(t)$ captures the underlying status of the system, which switches to a ``hot" state ($S(t) = 1$) immediately following an event and remains in the state for a random period of length $\tau$. If no further event occurs within $\tau$ time units after an event, the system returns to a ``regular" state ($S(t) = 0$). Note that, while the transitions into the "hot" state are triggered deterministically by events, the durations of ``hot" states are governed by the unobserved random variable $\tau$, making $S(t)$ a semi-Markov process. The model component $\lambda_0(t|\mathcal{H}(t),\pmb{Z})$ is the baseline event intensity during the ``regular" state (i.e. $S(t) = 0$); $\lambda_1(t|\mathcal{H}(t),\pmb{Z})$, the elevated intensity during the ``hot" state (i.e. $S(t) = 1$). In many practical situations, $\lambda_1(t|\mathcal{H}(t),\pmb{Z})$ is much larger than $\lambda_0(t|\mathcal{H}(t),\pmb{Z})$ for any fixed covariate $\pmb{Z}$, reflecting a temporary increase in event occurrence after an event occurs. There are examples for $\lambda_1(t|\mathcal{H}(t),\pmb{Z})$ 
smaller than $\lambda_0(t|\mathcal{H}(t),\pmb{Z})$.

We remark that, while the intensity function $\lambda(t|\mathcal{H}(t), \tau, \pmb{Z})$ in general depends on the full event history, the individual components $\lambda_0(t|\mathcal{H}(t),\pmb{Z})$ and $\lambda_1(t|\mathcal{H}(t),\pmb{Z})$ in the model may be specified independently from the history information $\mathcal{H}(t)$ with its influence summarized by the hidden binary process $S(t)$. The proposed model captures the potential self-exciting behavior of event occurrence via such a simple latent structure. As such, the model generalizes the conventional Hawkes process, offering greater flexibility for modeling event-driven dynamics where the excitation effect is transient and randomly lasting.

\subsection{Gap Time Distribution under the Proposed Model}
\label{sec:gap}

Denote the gap times between the consecutive event times by $Y_k= T_k - T_{k-1}$ for $k = 1, \dots$ with $T_0 = 0$. The $1$st gap time $Y_1$ is the time to the $1$st event (i.e. $T_1$ ) and its conditional hazard function is $\lambda_0(t|\pmb{Z})$ for $t>0$. The distribution of the $k$th gap time $Y_k$ for $k>1$, however, exhibits a piecewise structure: for $0<t<\tau-T_{k-1}$, the time within the duration of $\tau$ since the $(k-1)$th event, the event intensity is elevated to $\lambda_1(t|\mathcal{H}(t),\pmb{Z})$; after $\tau$ elapsed without a new event, the intensity returns to $\lambda_0(t|\mathcal{H}(t),\pmb{Z})$. 

To illustrate this feature, we consider a simple version of the model with constant $\lambda_0$ and $\lambda_1$ in the absence of covariates,
\begin{equation}
  \lambda(t|\mathcal{H}(t), \tau) = 
\lambda_1 S(t) +
\lambda_0\big[1-S(t)\big].
\label{casesimple}
\end{equation}
The waiting (gap) time until the first event follows the exponential distribution with the rate parameter$\lambda_0$. Since event intensities $\lambda_0, \lambda_1$ are independent of time, given the excitation duration $\tau$, the subsequent waiting times $Y_k$ for $k>1$ follow the same distribution with the density function $f(y|\tau)$ to be
\begin{equation}
f(y|\tau) = I(y\leq\tau)\lambda_1 \exp(-\lambda_1 y)+I(y>\tau)\lambda_0\exp(-(\lambda_1-\lambda_0)\tau)\exp(-\lambda_0 y).
\label{specialpdf}
\end{equation}
The distribution in (\ref{specialpdf}) has a piecewise structure involving the rates $\lambda_0, \lambda_1$. This distribution provides a change-point model: since the previous event and up until $y = \tau$, the waiting time follows an exponential distribution with rate of $\lambda_1$; after the change point $y = \tau$, the distribution shifts to an exponential distribution with rate  of $\lambda_0$ with a scaling factor $\exp(-(\lambda_1-\lambda_0)\tau)$. 

We refer to the distribution in (\ref{specialpdf}) as a piecewise distribution with a change point at $\tau$. Although our modeling is motivated by the analysis of soccer corner kicks, this change-point model can be applied to other fields. For instance, it may be useful for modeling patient survival times after surgery with a treatment threshold in the context of diseases like particular respiratory infections, which might pose high risks only for specific groups, such as children. The change-point model not only captures the potential of a mixture in survival times but also helps identify when the event rate shifts from one to the other.

\subsection{Specifications of Model Components}
\label{sec:spec}

\subsubsection{Random duration of the ``hot'' state}
\label{sec:tau}

The random duration time $\tau\geq 0$ is latent, one of the unique features of the proposed model. That yields the hot state indicator $S(\cdot)$ partly hidden: the beginning of the ``hot'' state is observable, whereas its end is not,
and the end of the ``regular'' state is observable, whereas its beginning is not, except at time $t=0$. The real data example presented in Section \ref{sec:analysis} assumes  $\tau$ follows a gamma distribution. We suggest an appropriate and flexible model be used for $\tau$'s distribution to capture the following two extreme cases. 

\begin{itemize}
    \item {\sl Case of $\tau \equiv 0$}.  The intensity function $\lambda(t|\mathcal{H}(t),\tau, \pmb{Z})$ reduces to $\lambda_0(t|\mathcal{H}(t),\pmb{Z})$, indicating no additional excitation from past events.
    
    \item {\sl Case of $\tau \gg 0$}. When the change-point $\tau$ is large relative to the length of the follow-up period $(0,E]$, the intensity approximates to $\lambda_1(t|\mathcal{H}(t),\pmb{Z})$, reflecting the event process remains in the high event intensity (i.e. ``hot'' state) after the initial event.
\end{itemize}

\subsubsection{Two component event intensity functions}
\label{sec:compo}

The proposed model (\ref{proposed}) bears considerable flexibility in specifying $\lambda_0(t|\mathcal{H}(t),\pmb{Z})$ and $\lambda_1(t|\mathcal{H}(t),\pmb{Z})$ to accommodate the special features in the application under consideration. One may choose each of them to be a commonly-used model for event intensity. Below are relevant special cases.
\begin{itemize}
    \item {\sl Extended Hawkes Process}. Specify $\lambda_1(t|\mathcal{H}(t),\pmb{Z})$ to be the intensity function
    of a self-exciting (i.e. Hawkes) process, say, $\lambda_1(t|\mathcal{H}(t),\pmb{Z}) 
    = \lambda_0(t|\pmb{Z}) + \sum_{k=0}^{N(t-)} w(t-T_k)$
    or $\lambda_1(t|\mathcal{H}(t),\pmb{Z}) 
    = \lambda_0(t|\pmb{Z}) + \sum_{k=0}^{N(t-)} w(t-T_{N(t-)})$ with $w(\cdot)$ a nonincreasing and nonnegative function.
    
    \item {\sl Mixture of Two Poisson Processes}. Consider $\lambda_0(t|\mathcal{H}(t),\pmb{Z})=\lambda_0(t|\pmb{Z})$
    and  $\lambda_1(t|\mathcal{H}(t),\pmb{Z})=\lambda_1(t|\pmb{Z})$. One may assume the Cox regression
    (i.e. Andersen-Gill) models for the two component intensity functions: $\lambda_0(t|\pmb{Z})=\lambda_{00}(t)\exp(\pmb{\beta}_0\pmb{Z})$ and $\lambda_1(t|\pmb{Z})=\lambda_{10}(t)\exp(\pmb{\beta}_1\pmb{Z})$. 
\end{itemize}

We illustrate the proposed model via the following four examples of the mixture of two Poisson processes in the
numerical studies of Section~3:
\begin{subequations}\label{examples}
\begin{itemize}
\item[] Model (a). With an additional parameter $\nu$ as the coefficient to the indicator of state ``hot'',
\begin{equation}
  \lambda(t|\mathcal{H}(t), \tau, \pmb{Z}) =
  \lambda_{00}\exp\{\pmb{\beta}\pmb{Z} + \nu S(t)\}.
  \label{case1}
\end{equation}

\item[] Model (b). Specifying the baseline function of $\lambda_0(t|\pmb{Z})$ up to an unknown parameter $\theta_{\lambda}$,
\begin{equation}
  \lambda(t|\mathcal{H}(t),\tau, \pmb{Z}) =
  \lambda_{0}(t;\theta_{\lambda})\exp\{\pmb{\beta}\pmb{Z}+\nu S(t)\}.
  \label{case2}
\end{equation}

\item[] Model (c). Keeping the baseline function of $\lambda_0(t|\pmb{Z})$ unspecified,
\begin{equation}
  \lambda(t|\mathcal{H}(t),\tau, \pmb{Z}) =
  \lambda_{00}(t)\exp\{\pmb{\beta}\pmb{Z}+\nu S(t)\}.
  \label{case3}
\end{equation}

\item[] Model (d). Allowing the covariate effects as well as baseline functions in the two states to be different,
\begin{equation}
  \lambda(t|\mathcal{H}(t),\tau, \pmb{Z}) =
 \lambda_{00}(t)\exp\{\pmb{\beta}_0\pmb{Z}\}\big[1-S(t))\big]+\lambda_{10}(t)\exp\{\pmb{\beta}_1\pmb{Z}\}S(t).
  \label{case4}
\end{equation}
\end{itemize}
\end{subequations}
The models in the above examples are nested and can enjoy a straightforward model checking. 

\subsection{Parameter Estimation}
\label{sec:est}

Suppose that $\{N_{i}(t), t \geq 0\}$ for $i=1,\ldots, n$ is a collection of independent realizations of the counting process under consideration. For the $i$th study unit, denote the associated covariates by $\pmb{Z}_i$; follow-up intervals, $(0, E_i]$; event occurrence times, $T_{ik}$ for $k=1,\ldots, K_i=N_i(E_i)$; the random ``hot'' state duration, $\tau_i$. This section presents the maximum likelihood estimation (MLE) precedure for estimating the model parameters of the mixture of two Poisson process model with the conditional intensity function specified as 
\begin{equation}
  \lambda(t|\mathcal{H}(t),\tau, \pmb{Z}) =
\lambda_{1}(t;\theta_{\lambda_1})\exp(\pmb{\beta}_1\pmb{Z})S(t)
+\lambda_{0}(t;\theta_{\lambda_0}) \exp(\pmb{\beta}_0\pmb{Z})\big[1-S(t)\big].
\label{case4procedure}
\end{equation}
The model above includes as special cases Examples 1-2 in the Section 
\ref{sec:compo}.
Further, we assume that the probability density function of $\tau$ is $f_{\tau}(\cdot;\theta_{\tau})$.

Let $\pmb{\theta}=(\theta_{\lambda_0},\theta_{\lambda_1}, \pmb{\beta}_0^{'}, \pmb{\beta}_1^{'}, \theta_{\tau})$. Conditional on $\tau_i$, the contribution of the study unit $i$ to the likelihood function is 
\begin{align}
    &L_{i}(\pmb{\theta}|\tau_{i})
     =  \prod_{k=1}^{K_i} \Big\{\lambda_{1}(T_{ik};\theta_{\lambda_1})\exp(\pmb{\beta}_1\pmb{Z}_i)S_i(T_{ik})
+\lambda_{0}(T_{ik};\theta_{\lambda_0}) \exp(\pmb{\beta}_0\pmb{Z}_i)\big[1-S_i(T_{ik})\big]\Big\} \nonumber\\
	& \times \exp\big\{-\int_0^{E_i} \lambda_{1}(t;\theta_{\lambda_1})\exp(\pmb{\beta}_1\pmb{Z}_i)S_i(t)
+\lambda_{0}(t;\theta_{\lambda_0}) \exp(\pmb{\beta}_0\pmb{Z}_i)\big[1-S_i(t)\big] dt\big\},
    \label{fulllikeli}
\end{align}
where $S_i(t) = I(t-T_{N_i(t-)} \leq \tau_{i})$. The likelihood function of $\pmb{\theta}$ with the observed
data is then
\begin{equation}
L(\pmb{\theta})=\prod_{i=1}^n \int L_{i}(\pmb{\theta}|\tau_{i}) f_{\tau}(\tau_i;\theta_{\tau}) d\tau_i.
\label{likel_obs}
\end{equation}
One may maximize the likelihood function (\ref{likel_obs}) to obtain the MLE of  $\pmb{\theta}$. That procedure can be
computationally intensive. Denote the log-transformed full data likelihood function by $l_{full}(\pmb{\theta};\pmb{\tau})=\sum_{i=1}^n \big[\log L_{i}(\pmb{\theta}|\tau_{i})  +\log f(\tau_i;\theta_{\tau})\big]$ with
$\pmb{\tau}=(\tau_1,\ldots,\tau_n)^{'}$. A commonly-used approach for computing the parameter MLE $\hat{\pmb{\theta}}$ is to adapt the MCEM algorithm as follows.

\begin{algorithm}[H]
\caption{MCEM Algorithm}
\label{algo-1}

1. Let the estimate at the $d$th iteration be $\pmb{\theta}^{(d)}$, with $\pmb{\theta}^{(0)}$ for the initial value.\;

\medskip
\textbf{2. Repeat}
\medskip

\Indp
2.1 MCE-step. Approximate 
$Q(\pmb{\theta},\pmb{\theta}^{(d)}) 
= E\!\left(l_{full}(\pmb{\theta};\pmb{\tau}) 
\mid \text{observed data}, \pmb{\theta}^{(d)}\right)$ 
using the sample mean$\tilde{Q}(\pmb{\theta},\pmb{\theta}^{(d)}) 
= \frac{1}{M}\sum_{m=1}^M 
l_{full}(\pmb{\theta};\pmb{\tau}^{(m)})$, where the $i$th component of $\pmb{\tau}^{(m)}$ is generated independently from the conditional distribution $[\tau_i \mid \text{observed data}, \pmb{\theta}^{(d)}]
\propto 
\big[L_i(\pmb{\theta}^{(d)} \mid \tau_i)\big]
f_{\tau}(\tau_i \mid \pmb{\theta}^{(d)}_{\tau}).$
\;

\medskip
2.2 M-step. Update the estimate of $\pmb{\theta}$ as
$\pmb{\theta}^{(d+1)}
= \operatorname*{arg\,max}_{\pmb{\theta}}
\tilde{Q}(\pmb{\theta},\pmb{\theta}^{(d)}).
$\;
\Indm
\medskip
\textbf{3. Iterate until} 
$\{\pmb{\theta}^{(d)}, d = 1,2,\ldots\}$ 
\textbf{converges}.

\end{algorithm}

One may obtain a consistent estimator of the standard error of the MLE $\hat{\pmb{\theta}}$
following \cite{louis1982finding}: the observed Fisher information matrix is now
\begin{equation*}
  \begin{split}
 E\big[-\frac{\partial^2l^{full}(\pmb{\theta})}{\partial\pmb{\theta}^2}|\text{observed data}, \pmb{\hat{\theta}}\big] - Var\big[\frac{\partial l^{full}(\pmb{\theta})}{\partial\pmb{\theta}}|\text{observed data}, \pmb{\hat{\theta}}\big].
   \end{split}
 \end{equation*}

When considering the mixture of two Poisson process model as in (\ref{case4procedure}) with unspecified baseline functions, such as in Models (c) or (d) of Section \ref{sec:compo}, each of the unspecified baseline line functions may be approximated by a piecewise constant or a spline (piecewise polynomial) function. The above estimation procedure can be adapted accordingly for estimating all the unknowns in the model.

\section{Numerical Studies}
\label{sec:analysis}

\subsection{Analyses of Chinese Super League 2019 Data}
\label{sec:data}

\subsubsection{Data and segment construction}
The 2019 Chinese Super League (CSL) season consisted of 16 teams, with each team playing every other team twice (home and away), yielding 240 scheduled matches. Event records for seven matches were unavailable, likely due to video or logging issues; these matches were excluded from the analysis and treated as missing completely at random. For each recorded match, corner kicks are available as time stamped recurrent events, which naturally motivates a counting process formulation.

A practical complication is that the corner kick process within a match is interrupted by ``terminal'' match events that reset the game context, such as goals and ends of halves. Moreover, the corner kick processes of the two teams are coupled in real time. To avoid modeling cross-team dependence directly, we conduct stratified analyses for the home and away teams, and we further partition each match into segments separated by terminal events.

Specifically, for match $i$ ($i=1,\ldots,233$), define a \emph{segment} as the interval that starts at a segment start event and ends at the next terminal event. In our application, segment start events include the start of a half and goals, while terminal events include goals and the ends of halves. Let $j=1,\ldots,J_i$ index the segments within match $i$. For a given team (home or away), let $\{N_{ij}(t),\,0\le t\le E_{ij}\}$ denote the segment-level corner kick counting process, where $N_{ij}(t)$ is the number of corner kicks observed up to segment time $t$ since the start of segment $j$, with $N_{ij}(0)=0$ and follow-up ending at $E_{ij}$. Let $\pmb{Z}_{ij}$ denote the vector of segment-level covariates (Table~\ref{table-covariates}). Across the 233 matches, a
total of 705 goals were observed, which, together with the 466
half-ending events, results in 1171 segments in the dataset.
\begin{center}
\textit{$<$Insert Table~\ref{table-covariates} here$>$}
\end{center}
The proposed model in Section~\ref{sec:mod} is fitted at the segment level. In particular, the latent ``hot state'' duration is allowed to vary across segments: $\tau_{ij}$ is treated as an unobserved random variable for each segment $(i,j)$. Across all segments, we assume $\tau_{ij}\overset{\text{i.i.d.}}{\sim}\Gamma(\text{shape},\text{rate})$, with the shape and rate parameters estimated from the data. Under this construction, each segment constitutes a new realization of the proposed event process, and terminal events (goals and half-time endings) induce a natural restart of the segment time scale.

\subsubsection{Model specifications and covariates}

We apply the four model specifications introduced in Section~\ref{sec:compo} (Models~(a)--(d)) to the CSL data. These examples form a nested sequence of models with increasing flexibility in the specification of the baseline intensity and covariate effects, ranging from a constant baseline (Model~(a)) to models allowing time-varying baselines and state-specific covariate effects (Models~(b)--(d)). All models are fitted using the MCEM procedure described in Section~\ref{sec:est}.

Exploratory analyses indicated systematic differences in the temporal patterns of corner kicks between the home and away teams, motivating the use of home/away-specific covariate sets. Candidate covariates and their definitions are summarized in Table~\ref{table-covariates}. Variable selection was performed using all-subset selection based on the Bayesian information criterion (BIC) under the model specification in Model~(a). This procedure was applied separately for the home and away teams. The same covariate sets were subsequently used for the more flexible specifications.

For the home team, the selected covariates include the score difference and an indicator of whether the segment starts immediately after a goal. For the away team, the selected covariates include an indicator for the second half, the score difference, the segment starting time, and an interaction term between the second half indicator and the segment starting time. These differences reflect distinct contextual dynamics faced by home and away teams.

The four model specifications differ in their treatment of the baseline intensity and state dependence:
\begin{itemize}
  \item \textbf{Model~(a):} A constant baseline intensity.
  
  \item \textbf{Model~(b):} A time-varying regular state baseline intensity specified through a parametric functional form. Specifically, the log baseline intensity is modeled as
  \[
  \log \lambda_{0}(t;\theta_{\lambda}) 
  = \theta_1 + \theta_2\big\{\exp(-\theta_3 t) - \exp(-\theta_4 t)\big\},
  \]
  which involves four unknown parameters and allows for flexible nonmonotone temporal patterns over a segment.
  
  \item \textbf{Model~(c):} A semiparametric specification in which the regular state baseline intensity is left unspecified and approximated by a piecewise constant function, as described in Section~\ref{sec:est}. Based on exploratory analyses of segment durations and corner kick timing, the cut points are chosen to be non-uniform, with higher resolution early in the segment where temporal variation is more pronounced. The regular state baseline is approximated using eight pieces with cut points at $0,\ 3,\ 6,\ 9,\ 12,\ 15,\ 27.5,\ \text{and } 40$ minutes.

  \item \textbf{Model~(d):} An extension of Model~(c) that additionally allows a separate baseline intensity for the hot state. The hot state baseline is also approximated by a piecewise constant function, using six pieces with cut points at $0,\ 5,\ 10,\ 15,\ 20,\ \text{and } 30 \text{ minutes}$.
  
\end{itemize}

\subsubsection{Parameter estimates and interpretation}
Table~\ref{exp123} reports the parameter estimates for Modelss~(a)--(c). Overall, the estimated regression effects, and the distribution of the latent hot state duration $\tau$ are highly consistent across these model specifications, indicating robustness to the choice of baseline intensity. 
\begin{center}
\textit{$<$Insert Table~\ref{exp123} here$>$}
\end{center}
For the home team, the estimated coefficients in the regular state suggest a lower corner kick rate when the team is leading and when a segment starts immediately after a goal. These effects are statistically significant and persist across all three model specifications. This
finding is consistent with existing evidence that leading teams tend to adopt less aggressive strategies, resulting in reduced attacking
pressure, as in \cite{silva2016analysis, guan2023parking}. For the away team, the estimates indicate higher corner kick rates in the second half of the match and lower rates when leading. In addition, the negative interaction between score difference and segment starting time suggests that the suppressive effect of leading becomes more pronounced as the match progresses. Compared with the home team, the remaining segment time plays a more important role in the away team’s corner kick intensity.

The estimated distribution of the latent hot state duration $\tau$ is stable across models and teams. For both home and away teams, the estimated mean duration of the hot state is close to two minutes, with comparable standard deviations. This finding suggests that, following a corner kick, any transient increase in the likelihood of a subsequent corner kick is typically confined to a short and well-defined time window. The similarity between home and away teams indicates that the temporal scale of such short-term excitation is largely team-invariant in this dataset. From a soccer perspective, one might expect such excitation effects to carryover a much shorter window (e.g., several seconds); however, the observed event times include player positioning and set-piece preparation, which effectively stretch the recorded time scale between corner kicks.

Table~\ref{exp4} presents the results for Model~(d), which allows for state-specific covariate effects. The estimated coefficients for the regular state are broadly consistent with those obtained in Models~(a)--(c). In contrast, none of the candidate covariates exhibit statistically significant effects in the hot state for either team. This suggests that, conditional on entering the hot state, the elevated corner kick rate is primarily driven by the excitation mechanism itself rather than by additional contextual covariates.
\begin{center}
\textit{$<$Insert Table~\ref{exp4} here$>$}
\end{center}
This absence of statistically significant covariate effects in the hot state is consistent with domain knowledge in soccer analytics. Following a corner kick, the play is highly stochastic, with limited opportunity for systematic tactical or contextual factors to exert a sustained influence. Given the short duration of the hot state and the relatively small number of quick follow-up corner kicks, detecting covariate effects would require larger sample sizes of soccer matches. Indeed, some estimated effects (e.g., score difference for the away team) exhibit non-negligible magnitudes and suggestive directional patterns, but the available data are insufficient to support statistical significance. For example, with richer data, it may be possible to consider and identify team-specific tendencies, such as a preference for short corner kicks, which may reduce the likelihood of immediate subsequent corners and are difficult to detect with the current sample size.

\subsubsection{Estimated baseline intensities}
\begin{figure}
  \centering
  \includegraphics[width=0.7\textwidth]{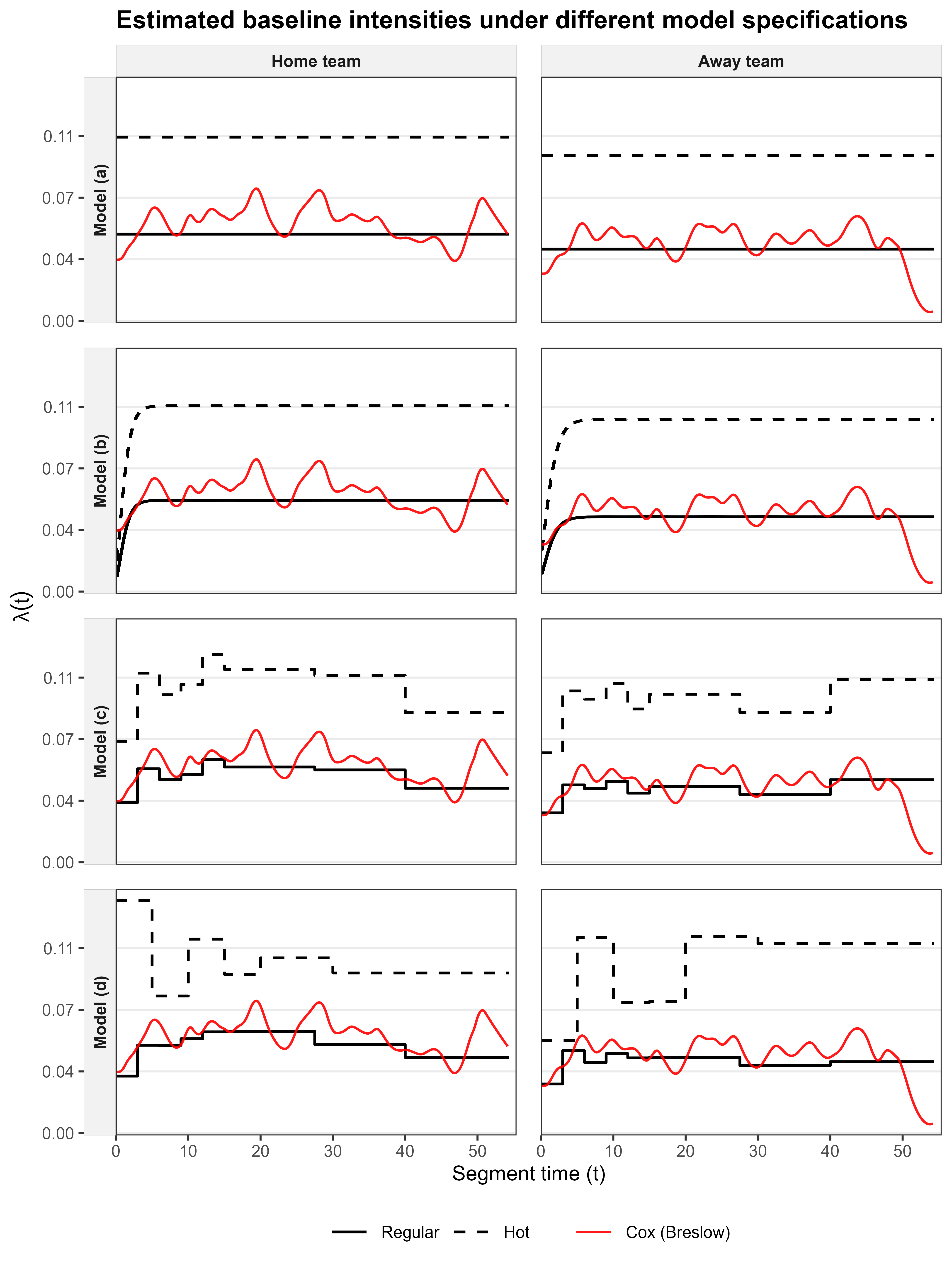}  
  \caption{Estimated baseline intensity functions for the home and away teams under four model specifications.
Each row corresponds to a different baseline specification (Models~(a)--(d)), and each column corresponds to team (home vs.\ away).
Within each panel, solid and dashed black curves represent the estimated baseline intensities under the regular and hot states, respectively.
The red curve shows a smoothed baseline intensity derived from a Andersen-Gill model using the Breslow estimator, included for reference.}
  \label{exp123fig}
\end{figure}

Figure~\ref{exp123fig} displays the estimated baseline intensity functions for the home and away teams under the four model specifications. Across models, the overall shapes and magnitudes of the estimated intensities are broadly comparable. Note that the time scale in Figure~\ref{exp123fig} corresponds to segment time rather than match or half time. In practice, only a small number of segments could extending to around 50 minutes, typically corresponding to an entire half with no goals plus extra time.
\begin{center}
\textit{$<$Insert Figure~\ref{exp123fig} here$>$}
\end{center}
At the same time, the results from Models~(b)--(d) clearly suggest that the baseline intensity varies over segment time. The fixed baseline assumption in Model~(a) appears insufficient to capture the temporal evolution of corner kick intensity within a segment. The time-varying specification in Model~(b) provides a flexible representation and closely tracks the nonparametric estimates obtained in Models~(c) and~(d), indicating that the chosen parametric form adequately captures the dominant temporal patterns in the data.

For reference, Figure~\ref{exp123fig} also includes a smoothed baseline intensity (red curves) estimated from a Andersen-Gill model \citep{andersen1982cox} using the Breslow estimator. The Cox baseline aligns closely with the estimated regular state baseline intensity from the proposed models, but is systematically slightly higher. This discrepancy reflects the fact that, without distinguishing between regular and hot states, the Cox model absorbs part of the transient excitation into a single baseline component, leading to a mild upward bias in the estimated regular state intensity and ignoring information about short-term clustering.

In Models~(a)--(c), the estimated hot state multiplicative effects $\hat{\nu}$ indicate that the corner kick intensity during the hot state is approximately twice that of the regular state. In contrast, Model~(d) allows the hot and regular baseline intensities to vary independently and does not impose a proportional structure. This observation suggests that the hot state intensity is not strictly proportional to the regular state intensity over time, highlighting a potential limitation of the proportionality assumption in Models~(a)--(c).

Overall, the estimated baseline intensities exhibit a common temporal pattern: corner kick intensity is low immediately after a segment begins, increases over the first several minutes, and then stabilizes. Throughout the segment, the home team generally displays a higher baseline corner kick intensity than the away team.

\subsubsection{An application: dynamic prediction illustration}
Beyond parameter estimation and inference on baseline intensities, the proposed modeling framework enables a range of applications that are not readily accessible through standard event-time models. For example, the models can be used to update predictions in real time as new events occur.

\begin{figure}[h!]
  \centering
  \includegraphics[width=0.7\textwidth]{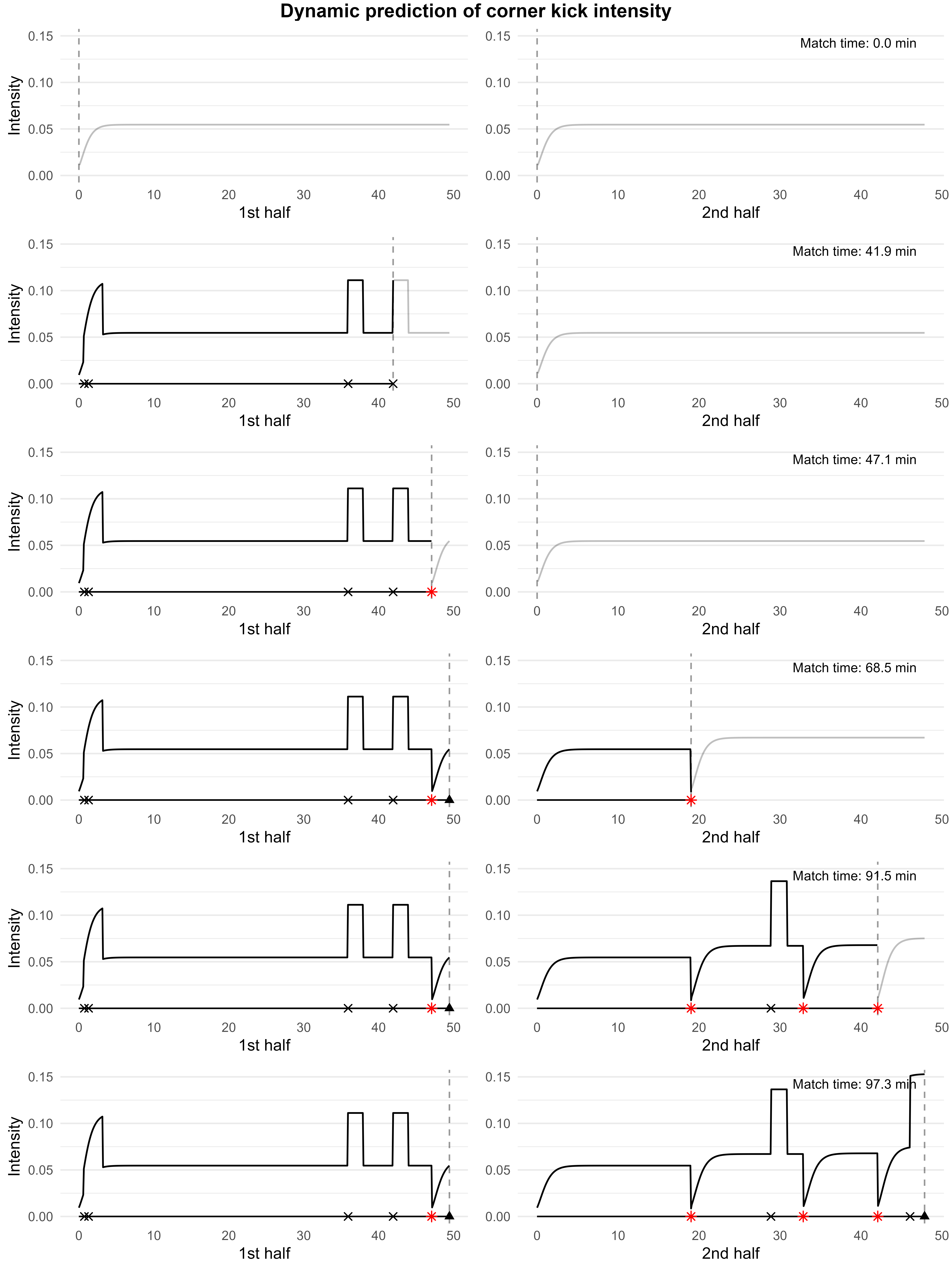}  
  \caption{Dynamic prediction of corner kick intensity (y-axis) for the home team in the soccer match
``Shenzhen vs Shanghai'' (Mar 1, 2019).
Six representative match times are shown.
For each row, the solid black curve displays the estimated intensity based on information observed up to the indicated time (vertical dashed line),
and the lighter extension shows the predicted future intensity conditional on the current state of the match.
Black crosses represent observed corner kicks, red stars represent goals, and black triangles indicate the end of each half.}
  \label{ingame}
\end{figure}

Figure~\ref{ingame} illustrates this capability through a dynamic prediction of the corner kick intensity for the home team in the match between Shenzhen and Shanghai on March~1,~2019, based on the fitted model in Model~(b). Six representative match times are shown. At each time point, the solid black curve represents the estimated intensity conditional on all information observed up to that time, while the lighter extension depicts the predicted future intensity for the remainder of the segment. The prediction is updated whenever a new event occurs, such as a corner kick or a goal.
\begin{center}
\textit{$<$Insert Figure~\ref{ingame} here$>$}
\end{center}
This example highlights how the model captures both the baseline temporal evolution of corner kick intensity within a segment and the short-term excitation following recent events. In particular, the predicted intensity increases immediately after a corner kick and subsequently decays as the system returns to the regular state, with the duration of this elevated period governed by the estimated distribution of the latent hot-state duration. For a given time point, one can also compute the implied mean time to the next corner kick, along with associated uncertainty intervals, which provides an additional quantitative check on the plausibility of the predicted intensities; further evidence on the reasonableness of these intensity estimates is provided in the subsequent simulation study.

Such dynamic predictions may be useful in a variety of applied settings, including real-time match analysis, tactical decision support, and betting market applications. More broadly, with the fitted model, it can be used to derive a variety of practically meaningful metrics within a match. For example, the estimated baseline intensity can be interpreted as a measure of a team’s underlying attacking pressure during different phases of play, while the hot-state parameters characterize short-term momentum and the team’s ability to sustain pressure. With additional covariates, the framework can also be extended to assess contextual effects such as formation changes, substitutions, or player-specific involvement, providing a principled way to link event-level dynamics with tactical and performance evaluation.

\subsection{Simulating Corner Kicks in Soccer Games}
\label{sec:sim}

This subsection introduces a simulation framework based on the proposed model in Section~\ref{sec:mod}. The purpose here is to present a practical algorithm for simulating event processes given fitted model parameters. Such an algorithm provides a convenient tool for generating synthetic event data that retain the key structural features implied by the model, and is useful for applied investigations and scenario exploration.

\subsubsection{Simulation algorithms}

We consider two simulation algorithms corresponding to different specifications of the baseline intensity. For the simpler setting considered in model~(\ref{casesimple}) in Section~\ref{sec:gap}, in which the regular and hot state intensities are constant over time, simulation can be carried out directly using the change-point gap time structure implied by the model. This procedure is summarized in Algorithm~\ref{alg:three}, and is closely related to the piecewise gap time distribution discussed in Section~\ref{sec:gap}.

\begin{algorithm}[H]
\caption{Simulate an event process under simplified model (\ref{casesimple})}
\textbf{Input: }$\lambda_0,\lambda_1,\tau,E$ 
\textbf{Initialize: }$t_0=0$, $n=0$, $h=0$ 
\While{$t_n < E$}{
  \uIf{$h=0$}{
    Generate $w \sim \mathrm{Exponential}(\lambda_0)$\;
    Set $t_{n+1}=t_n+w$\;
  }
  \Else($h=1$) {
    Generate $w_1 \sim \mathrm{Exponential}(\lambda_1)$\; 
    \uIf{$w_1 \le \tau$}{
      Set $t_{n+1}=t_n+w_1$\; \tcp{Event occurs within the hot window}
    }
    \Else{
      Generate $w_0 \sim \mathrm{Exponential}(\lambda_0)$\;
      Set $t_{n+1}=t_n+\tau+w_0$\; \tcp{No event during hot; switch to regular after $\tau$}
    }
  }
  \If{$t_{n+1} > E$}{\textbf{break}\;\tcp{Beyond the follow-up period}}
  $n=n+1$\; \tcp{Event count increases by 1}
  
  Set $h=1$\; \tcp{Immediately after an event, enter the hot state}
}
\textbf{return: }$\{t_k\}_{k=1,\ldots,n}$
\label{alg:three}
\end{algorithm}

For the more general setting with time-varying baseline intensities, direct gap time simulation is no longer available. In this case, we adopt a thinning-based approach \citep{lewis1979simulation} to generate event times from the proposed intensity function; the procedure is summarized in Algorithm~\ref{alg:tv}. For corner kick processes, this additional flexibility is well motivated, as our real data analysis suggests that the baseline intensity varies substantially over the course of a segment and is far from constant.

\begin{algorithm}[H]
\caption{Simulate an event process with time-varying baseline intensities (thinning)}
\textbf{Input: }$\lambda_0(t), \lambda_1(t), \tau, E,$ 
$\bar{\lambda} = \sup_{0\le t\le E}\max\{\lambda_0(t),\lambda_1(t)\}$ 
\textbf{Initialize: }$t_0=0$, $n=0$, $m=0$, $s_0=0$, $h=0$ 
\While{$s_m < E$}{
  Generate $w \sim \mathrm{Exponential}(\bar{\lambda})$\;
  Set $s_{m+1}=s_m+w$\;
  \If{$s_{m+1}>E$}{\textbf{break}\;\tcp{Beyond the follow-up period}} 
  \If{$s_{m+1}<h$}{
    Set $\lambda^\star=\lambda_1(s_{m+1})$ \tcp{Hot state when $s_{m+1}<h$}
  }
  \Else{
    Set $\lambda^\star=\lambda_0(s_{m+1})$ \tcp{Regular state when $s_{m+1}\ge h$}
  }
  Generate $D \sim \mathrm{Uniform}(0,1)$\;
  \If{$D\le \lambda^\star/\bar{\lambda}$}{
    $t_{n+1}=s_{m+1}$\;
    $n=n+1$\; \tcp{Event accepted}
    $h=t_n+\tau$\; \tcp{Enter hot state for $\tau$ time units}
  }
  $m=m+1$\;
}
\textbf{return: }$\{t_k\}_{k=1,\ldots,n}$\;
\label{alg:tv}
\end{algorithm}

\begin{figure}
    \centering
    \includegraphics[width=1\hsize]{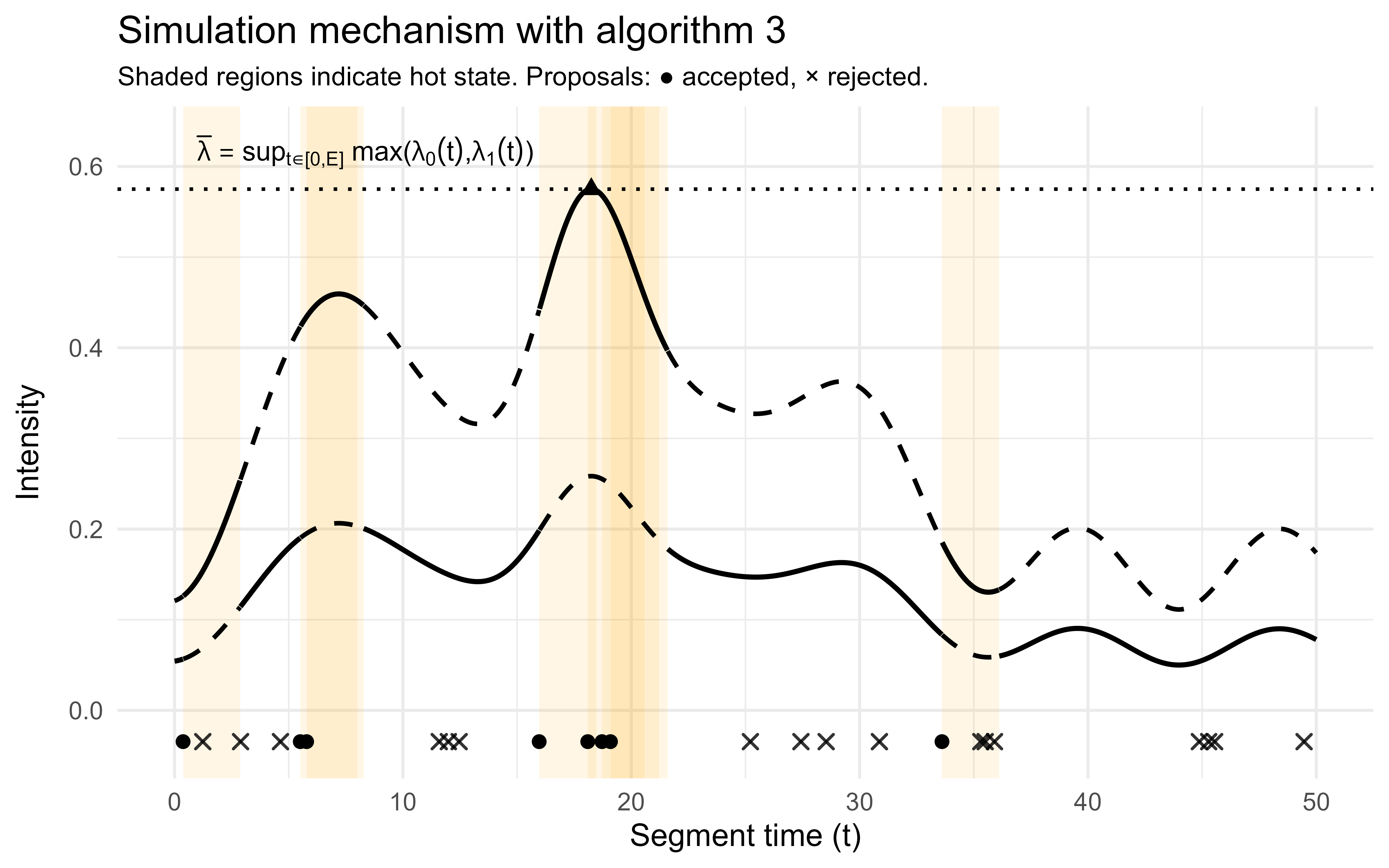}
\caption{Illustration of the thinning-based simulation algorithm for a time-varying intensity with a latent hot state. Solid and dashed curves indicate the active and inactive intensities, respectively; the solid curve corresponds to the effective intensity $\lambda^\star(t)$ used for thinning at time $t$. Shaded regions denote hot state windows following accepted events. Candidate events generated under the dominating rate are accepted with probability $\lambda^\star(t)/\bar{\lambda}$ (solid points) or rejected otherwise (crosses).}
\label{sim1}
\end{figure}

Figure~\ref{sim1} provides a schematic illustration of the simulation mechanism for a single segment. The regular state intensity $\lambda_0(t)$ and the corresponding hot state intensity $\lambda_1(t)$ are shown over segment time. Solid lines indicate the active intensity at a given time, while dashed lines represent the inactive intensity. Shaded regions correspond to hot state windows triggered by accepted events. Candidate event times are first generated from a homogeneous Poisson process with rate $\bar{\lambda}=\sup_{0\le t\le E}\max\{\lambda_0(t),\lambda_1(t)\}$, shown as points along the time axis. Each candidate time is then accepted as an event with probability $\lambda^\star(t)/\bar{\lambda}$, where $\lambda^\star(t)=\lambda_1(t)$ if the process is in the hot state and $\lambda^\star(t)=\lambda_0(t)$ otherwise. Accepted proposals are marked by solid points, while rejected proposals are indicated by crosses. This construction ensures that the accepted event times follow the desired time-varying intensity with transient excitation governed by the latent hot state duration.
\begin{center}
\textit{$<$Insert Figure~\ref{sim1} here$>$}
\end{center}

In the soccer corner kick application, these event-level simulation algorithms are embedded within a match-level structure. Each match is first partitioned into segments separated by terminal events such as goals and half-time boundaries, and corner kicks are simulated independently within each segment conditional on segment-specific covariates. The overall match-level simulation strategy follows the framework described in the supplementary material of \citet{peng2024}, while the segment-level corner kick processes are generated using the new simulation algorithms proposed here.

\subsubsection{Simulation results}

To demonstrate the simulation framework, we generated replicated seasons of corner kick data using the fitted model under Model~(b). Specifically, the simulation procedure was repeated 200 times, with each replicate consisting of 233 matches, matching the number of observed matches in the 2019 CSL season to facilitate direct comparison with the empirical data.

Terminal match events such as goals and half-time boundaries were also simulated following \citet{peng2024}. Match duration and goal times were generated from empirical distributions estimated from the CSL data, independently of the corner kick process. These terminal events partition each match into multiple segments, within which corner kicks were simulated using the event-level algorithms described above.

Overall, the simulated corner kick processes closely resemble the empirical CSL data in several simple and interpretable summaries. For both home and away teams, the simulated matches reproduce the observed average number of corner kicks per match (home: $5.21$ observed vs.\ $(4.95, 5.56)$ simulated; away: $4.72$ observed vs.\ $(4.36, 4.94)$ simulated), where the simulated values report the 2.5\% and 97.5\% empirical percentiles across the 200 replicated seasons. Similar agreement is observed for corner kick frequencies during the first 10 minutes of each half.

Overall, the simulated corner kick processes closely resemble the empirical CSL data in several simple and interpretable summaries. For both home and away teams, the simulated matches reproduce the observed average number of corner kicks per match (home: $5.21$ observed vs.\ $(4.95, 5.56)$ simulated; away: $4.72$ observed vs.\ $(4.36, 4.94)$ simulated), where the simulated values report the 2.5\% and 97.5\% empirical percentiles across the 200 replicated seasons, with comparable empirical spreads across matches. Another example is that the simulated data yield similar corner kick frequencies during the first 10 minutes of each half (home: $0.98$ observed vs.\ $(0.86, 1.15)$ simulated; away: $0.92$ observed vs.\ $(0.76, 0.99)$ simulated).

\begin{figure}
  \centering
    \includegraphics[width=1\hsize]{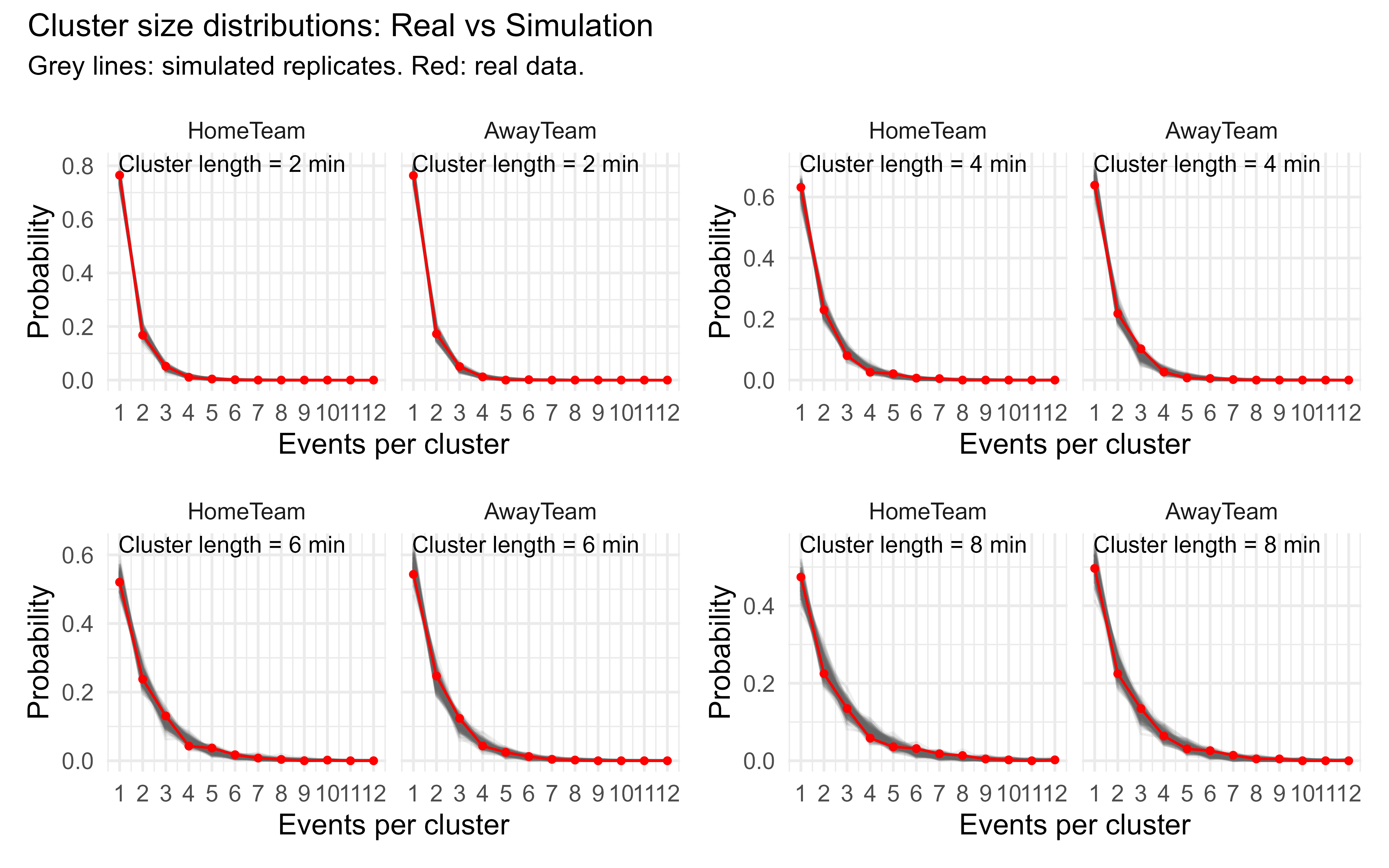}
\caption{Cluster size distributions for observed CSL matches and simulated matches. Cluster size is defined as the number of consecutive corner kicks occurring within a pre-specified time threshold. Panels correspond to different threshold durations, with separate curves for the home and away teams. The red curve represents the empirical probability mass function from the observed data, while the grey curves correspond to the 200 simulated seasons.}
\label{sim2}
\end{figure}

While the summaries above focus on overall frequency and broad temporal placement, these quantities can largely be derived directly from the estimated model parameters. A key practical advantage of the simulation algorithm is its ability to examine quantities of interest that are not easily obtainable in closed form from the parameter estimates alone. As an example, Figure~\ref{sim2} displays the cluster size distribution for both the observed CSL data and the simulated matches. The cluster size distribution summarizes the number of corner kicks occurring in rapid succession by grouping consecutive events within a pre-specified time threshold, providing a direct measure of short-term dependence in the event process.
\begin{center}
\textit{$<$Insert Figure~\ref{sim2} here$>$}
\end{center}
In Figure~\ref{sim2}, four panels corresponding to different cluster duration thresholds are shown. In each panel, the x-axis represents the number of events per cluster, while the y-axis shows the corresponding empirical probability mass function. Separate curves are presented for the home and away teams. The close agreement between the simulated and observed distributions indicates that the proposed simulation framework successfully reproduces the short-term clustering behavior observed in real matches. Beyond cluster size distributions, the proposed simulation framework can be used to examine a wide range of other event-level summaries that are difficult to derive analytically from fitted model parameters, such as run lengths of consecutive events, waiting time patterns conditional on recent history, or segment-level accumulation measures. These quantities can be explored through simulation, further highlighting the practical value of the proposed approach.

\section{DISCUSSION} 
\label{sec:dis}

This project proposed a new modeling framework for recurrent events using a self-exciting counting process with a random-memory excitation mechanism. The key feature of our model is that it defines the ``hot” state process $S(t)$ as a partly hidden semi-Markov process, which depends on both the event history and a latent random variable $\tau$. This setup allows the model to capture short-term excitation effects in a flexible and data-driven way. Unlike the traditional Hawkes process, which requires specifying a decaying excitation function, our model relies on a finite memory duration determined by $\tau$, making it easier to interpret and estimate. One special case of our model is similar to the carryover effect model in \cite{cciugcsar2012assessing}, but our use of a random $\tau$ allows for greater flexibility. Compared to the Markov-modulated Hawkes process of \cite{wu2022markov}, which uses a fully latent and Markovian state process, our model provides a more intuitive semi-Markov structure that directly ties the excitation to recent event history.

A second contribution of this work lies in the development of simulation algorithms tailored to the proposed model. Beyond parameter estimation, simulation plays a central role in understanding the implications of transient excitation for observable event patterns. We introduce practical algorithms for simulating event processes under both time-fixed and time-varying baseline intensities, including a thinning-based construction that accommodates the latent hot state mechanism. These algorithms enable the generation of synthetic event data that preserve the structural features implied by the fitted model. Importantly, they allow the investigation of quantities, such as short term clustering summaries, that are difficult to derive analytically from model parameters alone. As illustrated in the corner kick application, simulation provides a natural bridge between model specification and empirically interpretable event-level behavior.

In the context of soccer analytics, our model accounts for the dependence of corner kick events on past occurrences, unlike the independent event-time assumption in \cite{peng2024}. This dependence allows us to better capture the game’s flow and the clustering of events. Additionally, our model supports a Cox-type specification, where the baseline intensity can vary flexibly over time, making it well-suited to capture how corner kick rates change as the game progresses. This structure enables us to incorporate in-game covariates and better understand how match conditions affect the dynamics of corner kicks.

Several directions for future work arise from this study. Further investigation of the theoretical properties of the excitation duration distribution, including identifiability and sensitivity to distributional assumptions, would be valuable. On the computational side, extensions of the simulation framework to jointly model multiple interacting event processes, such as the home and away teams within a match, represent a natural next step, as real-game dynamics often involve inhibitory or competitive interactions.

Beyond soccer, the proposed framework is applicable to a wide range of recurrent event settings in which excitation is transient and event-triggered. Potential applications include epidemiology, criminology, finance, and reliability analysis, where short-term dependence plays an important role but is not well captured by long-memory excitation kernels. In such contexts, the combination of a parsimonious modeling structure and practical simulation tools provides a flexible foundation for both inference and exploratory analysis of complex event dynamics.

\section*{Acknowledgments}
This research is partially supported by the grants of Natural Sciences and Engineering Research Council of Canada (NSERC) to Hu and Swartz, and the CRT (Collaborative Research Team) in Sports Analytics of the Canadian Statistical Sciences Institute (CANSSI) led by Swartz.
The authors thank Daniel Stenz, former Technical Director of Shandong Luneng Taishan FC for providing the data discussed in this paper.

\section*{Declarations}
The authors declare no conflict of interest.

 \bibliographystyle{elsarticle-harv} 
 \bibliography{Manuscript}
 
\clearpage

\appendix
\clearpage

\section{Figures}

\clearpage

\section{Tables}

\begin{table}[!htb]
\centering
		\begin{tabular}{l|l}
			\toprule
			\textbf{Covariates} & \textbf{Description} \\
			\midrule
	    	\thead[l]{Indicator of 2nd half ($X_1$)}  &  0 if first half, 1 if second half \\
            \midrule
		      \thead[l]{Score difference ($X_2$)}        &   \makecell[l]{Positive $X_2$ means the team is leading,\\ Negative $X_2$ means the team is losing, \\
        Zero $X_2$ means a draw}\\
              \midrule
		      \thead[l]{Starts from goal ($X_3$)}        &  \makecell[l]{0 if segment starts from a actual start of the half, \\ 1 if the segment starts from a goal} \\
               \midrule
		      \thead[l]{Segment start time ($X_4$)}        &  \makecell[l]{A numerical variable indicates the starting time of the segment \\ since the start of the current half} \\
			\bottomrule
		\end{tabular}
\caption{Covariate list and descriptions}
\label{table-covariates}

\end{table}

\begin{table}[!htb]
\centering
\setlength{\tabcolsep}{1pt} 
\renewcommand{\arraystretch}{1.5} 
\scriptsize
\caption{Estimated $\pmb{\beta}$ from models (a), (b), and (c). Models are corresponding to first three rows in Figure (\ref{exp123fig}).}
\begin{threeparttable}
\begin{tabular}{r | c @{\hspace{8pt}} c | c @{\hspace{8pt}} c | c @{\hspace{8pt}} c }
    \toprule
    & \multicolumn{2}{c|}{\textbf{Model (a)}} & \multicolumn{2}{c|}{\textbf{Model (b)}} & \multicolumn{2}{c}{\textbf{Model (c)}}  \\
    \cmidrule(lr){2-3} \cmidrule(lr){4-5} \cmidrule(lr){6-7}  
    \textbf{Parameters}                          & \textbf{Home}  & \textbf{Away}  & \textbf{Home}  & \textbf{Away} & \textbf{Home}  & \textbf{Away} \\
    \midrule
    Indicator of 2nd half $X_1$ & $\cdot$             &  0.090 (.061)       & $\cdot$             &     0.083 (.061)         &  $\cdot$             & 0.091 (.061) \\
    Score difference $X_2$    & \pmb{-0.122 (.057)}   &  0.032 (.034)       & \pmb{-0.111 (.052)} &       0.033 (.034)             & \pmb{-0.116 (.056)}  & 0.036 (.034) \\
    Starts from goal $X_3$    & \pmb{-0.090 (.022)}   & $\cdot$             & \pmb{-0.094 (.023)} & $\cdot$            & \pmb{-0.091 (.023)} & $\cdot$ \\
    Segment start time $X_4$  & $\cdot$               & \pmb{-0.005 (.002)} & $\cdot$             &             -0.004 (.003)       & $\cdot$             & -0.004 (.003)  \\
    Interaction $X_4X_2$      & $\cdot$               & \pmb{-0.006 (.002)} & $\cdot$             &\pmb{-0.006 (.002)}         & $\cdot$             & \pmb{-0.006 (.002)}  \\
    \midrule
    $\tau$'s distribution & \tnote{1} $\Gamma$(8.76,4.43) & $\Gamma$(8.49,4.25) & $\Gamma$(9.46, 4.81) & $\Gamma$(7.09, 4.02) & $\Gamma$(9.12, 4.65) & $\Gamma$(9.18, 4.64)  \\
    $\tau$'s mean \& sd   &  1.98, 0.67 & 2.00, 0.69  & 1.97, 0.64 & 1.76, 0.66 & 1.96, 0.65 & 1.98, 0.65  \\
    \bottomrule
\end{tabular}
    \begin{tablenotes}\footnotesize
\item [1] $\Gamma$(shape, rate) denote gamma distribution with shape and rate parameters.
\end{tablenotes}
\end{threeparttable}
\label{exp123}
\end{table}

\begin{table}[!htb]
\centering
\setlength{\tabcolsep}{1pt}
\renewcommand{\arraystretch}{1.5}
\scriptsize
\caption{Estimated state-specific regression effects under model (d).}
\label{tab:exp4_beta}
\begin{threeparttable}
\begin{tabular}{r | c @{\hspace{8pt}} c }
    \toprule
    & \multicolumn{2}{c}{\textbf{Model (d)}} \\
    \cmidrule(lr){2-3} 
    \textbf{Parameters} & \textbf{Home}  & \textbf{Away} \\
        \midrule
        \multicolumn{3}{l}{For $\pmb{\beta}_0$ (regular state)} \\
    Indicator of 2nd half $X_1$ 
        & $\cdot$ 
        & 0.097 (.068) \\
    Score difference $X_2$        
        & \pmb{-0.104 (.026)} 
        & 0.019 (.038)  \\
    Starts from goal $X_3$   
        & \pmb{-0.133 (.065)} 
        & $\cdot$  \\
    Segment start time $X_4$   
        & $\cdot$ 
        & -0.005 (.003)  \\
    Interaction $X_4X_2$    
        & $\cdot$ 
        & \pmb{-0.006 (.002)}  \\
     \midrule
    \multicolumn{3}{l}{For $\pmb{\beta}_1$ (hot state)} \\
    Indicator of 2nd half $X_1$ 
        & $\cdot$ 
        & 0.077 (.156) \\
    Score difference $X_2$    
        & -0.012 (.066) 
        & 0.120 (.095) \\
    Starts from goal $X_3$  
        & -0.012 (.153) 
        & $\cdot$ \\
    Segment start time $X_4$ 
        & $\cdot$ 
        & 0.004 (.007) \\
    Interaction $X_4X_2$    
        & $\cdot$ 
        & -0.006 (.005) \\
    \midrule
    $\tau$'s distribution 
        & \tnote{1} $\Gamma$(9.18, 4.60) &  $\Gamma$(9.06, 4.57)\\
    $\tau$'s mean \& sd  
        & 1.99, 0.66 & 1.98, 0.66 \\
    \bottomrule
\end{tabular}
\begin{tablenotes}\footnotesize
\item[1] $\Gamma$(shape, rate) denotes a Gamma distribution with shape and rate parameters.
\end{tablenotes}
\end{threeparttable}
\label{exp4}
\end{table}

\end{document}